\RequirePackage{fix-cm}
\documentclass[conference,10pt]{IEEEtran}
\IEEEoverridecommandlockouts
\usepackage{cite}
\usepackage{amsmath,amssymb}
\usepackage{graphicx}
\usepackage{booktabs}
\usepackage{tabularx}
\usepackage{multirow}
\usepackage{xcolor}
\usepackage{colortbl}
\usepackage{url}
\usepackage{tikz}
\usepackage{afterpage}
\usepackage{enumitem}
\usepackage{capt-of}
\usepackage{pifont}
\usetikzlibrary{positioning,arrows.meta,fit,calc,shapes.geometric,patterns}
\usepackage[hidelinks]{hyperref}
\definecolor{vblue}{HTML}{7799D2}\definecolor{vblueL}{HTML}{D9E2F3}\definecolor{vblueM}{HTML}{C4D2EC}
\definecolor{vred}{HTML}{FF5959}\definecolor{vredL}{HTML}{FFDADA}
\definecolor{vorange}{HTML}{FEA932}\definecolor{vorangeL}{HTML}{FFECD2}
\definecolor{vpurple}{HTML}{9A8BC1}\definecolor{vpurpleL}{HTML}{E9E5F1}
\definecolor{vteal}{HTML}{68CCC6}\definecolor{vtealL}{HTML}{D2F0EF}
\definecolor{panelline}{HTML}{CFD5DC}
\definecolor{ink}{HTML}{1E2329}
\definecolor{inkmuted}{HTML}{5F6873}
\definecolor{tool}{HTML}{46515C}
\definecolor{model}{HTML}{9A8BC1}
\definecolor{accent}{HTML}{547DB8}
\definecolor{lane}{HTML}{F7F8FA}
\definecolor{blockfill}{HTML}{D9E2F3}
\definecolor{codefill}{HTML}{F7F8FA}
\definecolor{pass}{HTML}{7799D2}
\definecolor{fail}{HTML}{E3E7EC}
\definecolor{accentB}{HTML}{E5484D}
\definecolor{errfill}{HTML}{FFDADA}

\newcommand{\bench}{SDCBench-X}
\newcommand{\sdcoder}{SDCoder}
\newcommand{\step}[1]{\ding{\the\numexpr201+#1\relax}}

\begin{document}

\title{SDCoder: LLM-Based Timing Constraint Generation via RL and Type-Routed Reference Injection}

\author{\IEEEauthorblockN{Qingyun Zou, Bin Gao, Chunlin Tian, Wenju Zhao, Mingjia Fan, Zhou Wang,\\ Weng-Fai Wong, Bingsheng He, and Tulika Mitra}
\IEEEauthorblockA{National University of Singapore}}

\maketitle

\begin{abstract}
Synopsys Design Constraints (SDC) scripts define the timing requirements used in synthesis and static timing analysis. Generating correct SDC scripts from specifications and RTL is challenging because a script may execute successfully while still missing required constraints or targeting incorrect design objects. We present SDCoder, a locally deployable 14B model for automatic SDC generation. We first build 61,826 verified training examples and train the model using supervised fine-tuning followed by reinforcement learning (GRPO) with OpenSTA-based rewards. However, we find that GRPO struggles with constraint types that rarely appear correctly in sampled outputs, particularly multicycle paths. To address this limitation, we measure recall for each constraint type and identify those that remain poorly learned. We then fine-tune the model on complete reference scripts containing these types, together with replay data to preserve previously learned capabilities. We also introduce SDCBench-X, a 140-task benchmark that evaluates generated scripts by comparing their resulting constraint states with references. SDCoder passes 87 of 100 general tasks and 39 of 40 rare-type tasks, compared with 84 and 4 for GRPO alone and 72 and 22 for GPT-6.1-sol. On the existing SDCBench, SDCoder matches GPT-6.1-sol at pass@1, passing 19 of 22 tasks.
\end{abstract}

\section{Introduction}
\label{sec:intro}

Timing constraints determine what synthesis and static timing analysis tools check. Engineers express these requirements in Synopsys Design Constraints (SDC) scripts, which define clocks, I/O delays, operating modes, and timing exceptions~\cite{gangadharan2013constraining}.
Writing correct SDC scripts requires translating design specifications and register-transfer-level (RTL) clocking into precise constraints on design objects (Fig.~\ref{fig:example}).
For example, a generated clock needs the correct source and edge ratio, a multicycle path needs exact endpoints, and adding an input delay for another clock must preserve the existing delay.
Errors can be difficult to detect: a missing clock may leave registers unconstrained, while an incorrect false path may exclude valid paths from timing analysis~\cite{du1989false,coudert2010false}.
Scripts with either error can load without errors because successful execution does not guarantee agreement with the specification.

Large language models (LLMs) offer a way to automate SDC generation, but design confidentiality makes local deployment important.
Existing approaches such as LLM4SDC~\cite{han2026llm4sdc} rely on hosted general-purpose models, which may require sending unreleased specifications and RTL to external services.
Although LLMs have shown promise in RTL and electronic design automation (EDA) script generation~\cite{thakur2023benchmarking,liu2023verilogeval,lu2024rtllm,liu2024openllmrtl}, existing open models struggle with SDC generation.
In our evaluation, an unmodified 14B code model and two RTL models trained with reinforcement learning (RL) achieve no semantic passes at pass@1 on the 22 SDCBench~\cite{han2026llm4sdc} designs (\S\ref{sec:exp}).
This motivates a dedicated local SDC generator.

\begin{figure}[t]
\centering
\includegraphics[width=\columnwidth]{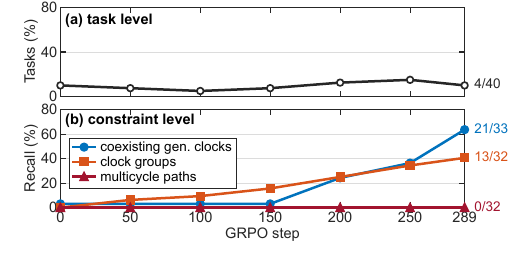}\par\vspace{-8pt}
\caption{Rare-type task passes and per-type recall during extra GRPO from R4 (Table~\ref{tab:reach}).}
\label{fig:insight}
\vspace{-8pt}
\end{figure}

Supervised fine-tuning (SFT) followed by RL with verifiable rewards, often using Group Relative Policy Optimization (GRPO)~\cite{shao2024deepseekmath}, has improved RTL generation~\cite{wang2025verireason,zhu2025codevr1,teng2025verirl}.
However, RL depends on the candidates sampled by the model: if none contains a correct instance of a required constraint, there is no correct instance of that constraint to reinforce~\cite{yue2025doesrl,wu2025invisible,ma2026relift,yan2025luffy}.

\afterpage{\begin{table}[t]
\includegraphics[width=\columnwidth]{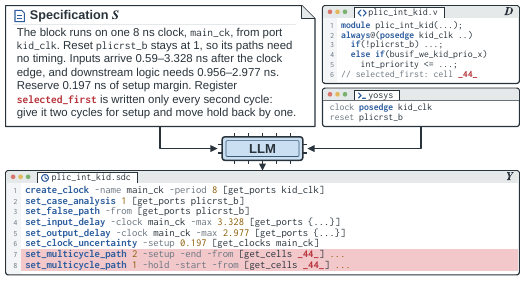}\par\vspace{-2pt}
\captionof{figure}{Task example (abbreviated).}
\label{fig:example}
\vspace{-10pt}
\end{table}}

Our key observation is that, for SDC, the verifier itself decomposes correctness. Unlike an RTL testbench, which returns one pass or fail per design, OpenSTA~\cite{opensta} executes a script and exports its constraint state, so each constraint can be compared exactly with the reference, without a learned reward model.
This lets us measure, per constraint type, which constraints GRPO learns and which remain difficult.
During one GRPO run from R4, evaluated on the 40 rare-type tasks (\S\ref{sec:bench}), the number of passed tasks hovers around 4 because a task passes only if every constraint is correct (Fig.~\ref{fig:insight}(a)): on one interrupt-controller task, the fraction of reference constraint facts reproduced correctly rises from 22\% to 89\%, yet the task still fails on its multicycle path.
Per type (Fig.~\ref{fig:insight}(b)), RL improves types that are rarely but sometimes sampled correctly: correct clock-group statements increase from 0 to 13 of 32.
In contrast, multicycle paths remain at 0 of 32 even with longer training or a 5$\times$ learning rate. 

To address this limitation, we propose \sdcoder{}, trained with \emph{type-routed reference injection}, which targets constraint types missed after RL (Fig.~\ref{fig:overview}).
We first construct verified training data (\S\ref{sec:coaug}) and train a local model using SFT and GRPO with OpenSTA-based rewards.
After GRPO, we measure recall separately for each constraint type and identify types with recall below 0.05 as \emph{missed types}.
We then fine-tune the model on complete reference scripts containing these missed types to improve their generation, while replaying earlier training data to preserve previously learned capabilities (\S\ref{sec:route}).
Our contributions are:
\begin{itemize}[leftmargin=0pt]
    \item \textbf{Type-level reachability of RL.}
    GRPO improves only the constraint types it sometimes samples correctly; we measure this per type from the exported state and add complete reference scripts only for missed types (\S\ref{sec:reach}, \S\ref{sec:routing}).

    \item \textbf{Training data, benchmark, and an open model.}
    To our knowledge, we build the first training set dedicated to SDC generation, 61,826 examples checked for agreement between specification and SDC (\S\ref{sec:coaug}); the 140-task \bench{} benchmark, evaluated by exported constraint state (\S\ref{sec:bench}); and \sdcoder{}, the first open 14B model trained for SDC generation.

    \item \textbf{Evaluation.}
    \sdcoder{} passes 87 of the 100 general \bench{} tasks, against 72 for GPT-6.1-sol, and matches GPT-6.1-sol on SDCBench at pass@1 (\S\ref{sec:main}).
\end{itemize}

\section{Background and Related Work}
\label{sec:related}

\noindent\textbf{Timing constraints.}
STA checks every timing path between sequential elements and ports against the timing environment defined in SDC~\cite{gangadharan2013constraining}.
Clocks (\texttt{create\_clock}) set periods and waveforms at ports or pins; generated clocks derive from a master clock through dividers or multiplexers.
Input and output delays model signal arrival and required times outside the block relative to a clock, while clock uncertainty and latency add margins.
Timing exceptions modify the default single-cycle check: false paths remove paths from analysis, multicycle paths allow several cycles, and clock groups mark clock pairs whose interactions are not timed.
Case analysis fixes mode signals to constants.

\noindent\textbf{LLMs for chip design.}
LLMs have been applied to RTL generation and EDA automation~\cite{thakur2023benchmarking,liu2023verilogeval,lu2024rtllm,liu2024openllmrtl,he2023chateda,zhao2025mage}.
RTL generators use GRPO~\cite{shao2024deepseekmath,guo2025deepseekr1} with testbench-based rewards~\cite{wang2025verireason,zhu2025codevr1,teng2025verirl}, and CraftRTL~\cite{liu2025craftrtl} generates repair data for checkpoint error categories.
LLM4SDC~\cite{han2026llm4sdc} uses a hosted general-purpose model with multi-agent orchestration and inference-time checks.
Its false-path discovery simulates each design with a functional testbench, but such testbenches are rare and cannot score thousands of RL training examples; our rewards need only a reference SDC.
Its SDCBench scores, from command acceptance and expert review, are not directly comparable with our task-level pass metrics.
We instead train a local model targeting constraint types missed after RL.

\section{Design}
\label{sec:method}
\label{sec:design}
\begin{figure*}[t]
\centering
\includegraphics[width=\textwidth]{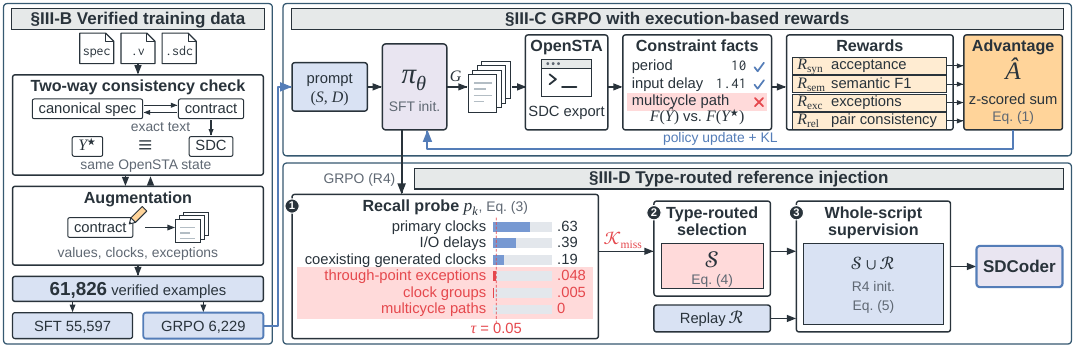}\par\vspace{-2pt}
\caption{Framework overview. Init.: initialization; $\mathcal{K}_{\mathrm{miss}}$: diagnosed missed types.}
\label{fig:overview}
\vspace{-8pt}
\end{figure*}

Our design builds on the ability to evaluate SDC correctness through the constraint state exported by a timing engine. As shown in Fig.~\ref{fig:overview}, \sdcoder{} follows a three-stage training pipeline.
First, we construct verified specification--RTL--SDC examples to provide reliable supervision (\S\ref{sec:coaug}).
Second, we train the model with SFT and execution-based GRPO, using constraint-state agreement as reward feedback (\S\ref{sec:grpo}).
However, GRPO struggles with constraint types that rarely appear correctly in sampled outputs.
We then diagnose missed types using constraint-level recall and inject complete reference scripts containing these types, together with replay data to preserve previously learned capabilities (\S\ref{sec:route}).
Together, these stages use execution-based verification to guide training and repair the remaining weaknesses of the model.

\subsection{Problem Setting}
\label{sec:problem}

\noindent\textbf{Task.}
Given a timing specification $S$ and RTL design $D$, the goal is to generate an SDC script $Y$ for the synthesized top module (Fig.~\ref{fig:example}). The model receives the specification, the RTL files, and a Yosys~\cite{wolf2013yosys} clock and reset summary.

\noindent\textbf{Constraint-state agreement.}
An SDC script can execute successfully while omitting required constraints, targeting incorrect design objects, or introducing unintended timing exceptions. We therefore check its resulting constraint state rather than command acceptance alone.
We synthesize $D$ using Nangate45~\cite{nangate45}, execute $Y$ in OpenSTA~\cite{opensta}, and export the resulting constraints with OpenSTA's SDC writer. Let $F(Y)$ denote the set of canonical facts representing exported clocks, periods, waveforms, sources, delays, exceptions, and case-analysis values, together with their design objects. A candidate $\hat{Y}$ semantically matches reference $Y^\star$ if $F(\hat{Y})=F(Y^\star)$, with numerical tolerance $10^{-6}$ in each field's units. This comparison is independent of equivalent Tcl notation.
For each constraint type $k$, let $F_k(Y)\subseteq F(Y)$ contain its facts. A type corresponds to an SDC command, optionally refined by an option, as in \emph{through-point exceptions} (\texttt{-through}) and \emph{coexisting generated clocks} (\texttt{-add}).

\subsection{Verified Training Data}
\label{sec:coaug}
\label{sec:admission}
\label{sec:sft}

Execution-based training requires trustworthy references: an incorrect SDC can load cleanly and still reward incorrect behavior. We therefore check specification--SDC agreement before including examples in training.

\noindent\textbf{Two-way consistency check.}
We collect specification, RTL, and SDC triples from public repositories and synthesize the RTL with Yosys. Every example must load cleanly in OpenSTA and match its stated constraints. Examples that also have a \emph{timing contract}, a typed representation of their timing requirements, undergo two further consistency checks.
First, parsing a canonical specification must produce the contract, and rendering that contract must recover the canonical specification exactly. Second, an SDC generated from the contract must produce the same exported OpenSTA state as the reference SDC on the synthesized netlist. Examples failing either check are discarded. The natural-language specification given to the model is then generated from the verified canonical specification. The exact text round-trip check applies only to the canonical representation, not to the natural-language text. These checks are selective: their latest round accepted 1,028 of 11,901 candidate designs, and augmentation rejected another 726 invalid variants.

\noindent\textbf{Constraint-aware augmentation.}
Public designs provide limited coverage of uncommon constraints, particularly generated clocks and clock groups. To broaden coverage while preserving consistency, we modify verified contracts and regenerate their canonical specifications and reference SDCs.
Augmentations vary timing values and units, object names, clock edges, virtual clocks, exceptions, clock groups, and specification templates. We also add clock dividers or multiplexers around unchanged blocks and check their clock behavior through Yosys simulation. The natural-language specifications are generated from the resulting canonical specifications, and all variants must pass consistency checks.

\noindent\textbf{Data split.}
The final dataset contains 61,826 verified examples, including 11,883 originals. To limit leakage across related variants, we group examples by source component and merge components sharing a structurally identical RTL module. We reserve 6,229 examples for GRPO and use the other 55,597 for SFT. No benchmark module appears in training.

\subsection{GRPO with Execution-Based Rewards}
\label{sec:pairs}
\label{sec:grpo}
\label{sec:rtl}

SFT teaches the model to imitate verified scripts but does not directly optimize the correctness of their resulting constraint states. We then apply GRPO to the SFT model, using OpenSTA to evaluate $G$ sampled scripts per prompt. Fig.~\ref{fig:overview} shows the reward path for prompts with reference SDCs.

\noindent\textbf{Rewards with a reference.}
For prompts with reference SDCs, each candidate $\hat{Y}$ is run on the synthesized netlist, and its exported facts are compared with $F(Y^\star)$. We use four rewards:

\begin{itemize}[leftmargin=*]
    \item \emph{Command acceptance} $R_{\mathrm{syn}}$ is the number of commands accepted without errors or warnings, divided by the larger of the candidate and reference command counts.

    \item \emph{Semantic F1} $R_{\mathrm{sem}}$ is the macro-average of fact-level F1 scores over SDC commands present in either state.

    \item \emph{Exception reward} $R_{\mathrm{exc}}$ penalizes extra facts more heavily than missing facts for false paths, multicycle paths, clock groups, minimum and maximum path delays, timing-arc disabling, and case analysis. This discourages unrequested constraints that may suppress required timing checks.

    \item \emph{Pair consistency} $R_{\mathrm{rel}}$ rewards outputs that preserve the relation within a prompt pair, an original prompt and a variant derived by a contract edit such as scaling all timing values by 0.5. It helps detect inconsistencies that may not prevent a script from loading.
\end{itemize}

\noindent\textbf{Rewards without a reference.}
We also train on synthetic clock and reset generators (CRGs) without reference SDCs. Yosys simulation records the registers reached by each clock and their clock ratios. The \emph{RTL clocking reward} $R_{\mathrm{rtl}}$ checks register-clock coverage, simulated clock ratios, and the specified input period. An \emph{RTL rule reward} penalizes false paths and clock groups rejected by RTL rules, duplicate clocks, and clocks on non-clock objects. These replace $R_{\mathrm{sem}}$ and $R_{\mathrm{exc}}$ for reference-free prompts, without evaluating I/O timing.

\noindent\textbf{Policy update.}
Rewards are standardized within the $G$ candidates of each prompt ($z_G$) and combined into the advantage
\begin{equation}
\label{eq:grpo-advantage}
\hat{A}
=0.25\,z_G(R_{\mathrm{syn}})
+z_G(R_{\mathrm{sem}})
+z_G(R_{\mathrm{exc}})
+0.5\,z_G(R_{\mathrm{rel}}).
\end{equation}
For reference-free prompts, RTL rewards replace $R_{\mathrm{sem}}$ and $R_{\mathrm{exc}}$. A truncated script, one whose constraint state cannot be exported, or one with no accepted command receives advantage $-1$. We update the policy with a policy-gradient objective with a Kullback--Leibler (KL) penalty over five GRPO rounds (R1--R5) with increasing data coverage (\S\ref{sec:exp}).

\subsection{Type-Routed Reference Injection}
\label{sec:route}

Type-routed reference injection targets constraint types that the GRPO model rarely generates correctly. GRPO relies on correct behaviors appearing in sampled outputs. Let $q_k$ be the probability that a sampled script contains a correct type-$k$ statement for a prompt requiring that type. The probability that at least one of $G$ candidates does so is
\begin{equation}
\label{eq:group}
1-(1-q_k)^G \approx Gq_k
\quad (q_k\ll 1).
\end{equation}
When $q_k$ is very small, ordinary GRPO groups rarely contain a correct instance to reinforce.

\noindent\textbf{\step{1}~Constraint-type recall probe.}
After GRPO, we run R4 (unless stated otherwise) greedily on the examples $x\in\mathcal{P}$ of the GRPO training pool (\S\ref{sec:coaug}) that have reference SDCs. For each constraint type $k$ present in the references, we compute
\begin{equation}
\label{eq:recall}
p_k =
\frac{
\sum_{x\in\mathcal{P}}
|F_k(\hat{Y}_x)\cap F_k(Y^\star_x)|
}{
\sum_{x\in\mathcal{P}}
|F_k(Y^\star_x)|
}.
\end{equation}
This measures the fraction of reference facts of type $k$ reproduced by the model. Types with $p_k<\tau=0.05$ form the set of \emph{missed types}, $\mathcal{K}_{\mathrm{miss}}$. If $p_k$ is read as a sampling rate $q_k$, a missed type has $q_k<\tau=0.05$; with $G=4$, $Gq_k<0.2$, fewer than one correct candidate per five groups on average (Eq.~\ref{eq:group}); any threshold between 0.048 and 0.19 selects the same set (\step{1} in Fig.~\ref{fig:overview}).
A missed type may be omitted or generated incorrectly; low greedy recall does not imply zero sampling probability or that the type is unlearnable.

\noindent\textbf{\step{2}~Type-routed selection.}
We select verified examples whose reference scripts contain at least one missed type:
\begin{equation}
\label{eq:select}
\mathcal{S} =
\left\{
x\in\mathcal{P}:
F_k(Y^\star_x)\neq\emptyset
\text{ for some } k\in\mathcal{K}_{\mathrm{miss}}
\right\}.
\end{equation}
This directs supervision toward constraint types that remain weak after GRPO. By contrast, \emph{task-failure selection}, inspired by ReLIFT~\cite{ma2026relift}, selects GRPO-pool examples whenever the R4 greedy output fails the exported-state match, $F(\hat{Y}_x)\neq F(Y^\star_x)$, regardless of failure type.

\noindent\textbf{\step{3}~Whole-script supervision.}
We resume fine-tuning from the GRPO checkpoint using complete reference scripts for the selected examples, including clock-wrapper variants. We do not restrict the loss to missed-type statements: correct SDC generation requires both producing requested constraints and avoiding unrequested ones. Whole-script supervision gives the targeted statements with their surrounding constraints and shows when other statements should be absent.

We combine the selected examples $\mathcal{S}$ with replay data $\mathcal{R}$. Let $\tilde{Y}_x$ denote the training target: the reference SDC for examples with reference SDCs, or a GRPO candidate that passes all checks with full reward for reference-free replay examples. Injection minimizes
\begin{equation}
\label{eq:inj-loss}
\mathcal{L}_{\mathrm{inj}}
=
-\sum_{x\in\mathcal{S}\cup\mathcal{R}}
\log\pi_\theta(\tilde{Y}_x\mid x).
\end{equation}

\noindent\textbf{Replay for retention.}
Focusing on missed types can degrade previously learned behavior. We therefore mix injection examples with replay data from earlier training. \emph{GRPO-data replay} includes examples from GRPO training with reference SDCs and reference-free CRG prompts paired with GRPO candidates that passed all checks with full reward. \emph{SFT-data replay} adds a uniform sample of the original SFT training data. Both use next-token loss. We also evaluate a KL penalty to the starting model as an alternative retention mechanism~\cite{marek2026forgetting}.
\subsection{The \bench{} Benchmark}
\label{sec:bench}

We evaluate both general SDC generation and the ability to produce uncommon constraint types that RL may miss. Existing SDCBench~\cite{han2026llm4sdc} covers these requirements only partially. Its 22 specifications primarily describe designs functionally; none specifies input or output delays, only three request multicycle paths or clock groups, and 21 already provide example clock definitions. Without reference SDCs, its semantic checks cover RTL clocking rather than the full constraint state (\S\ref{sec:exp}). We therefore construct \bench{}, a 140-task benchmark with reference SDCs evaluated through exported constraint-state agreement. It contains 100 general tasks and 40 rare-type tasks.
The \emph{general tasks} use unchanged public RTL from 65 repositories and reflect common block-level timing requirements: 91 specify input and output delays, 88 clock uncertainty, 67 mode control, and 61 static-control false paths. They also cover less frequent constraints, including falling-edge I/O, source latency, virtual and generated clocks, clock exclusivity, multicycle paths, and clock-domain crossings. These tasks contain 13 rare constraint statements: 6 multicycle paths, 3 through-point exceptions, and 4 coexisting generated clocks.
The 40 \emph{rare-type tasks} exercise scenarios involving through-point exceptions, exceptions targeting pins, clock multiplexers, case analysis on bus signals, multicycle paths, and coexisting generated clocks. These tasks were fixed before the missed-type diagnosis (\S\ref{sec:reach}). Their RTL comes from 12 public repositories with source directories disjoint from training; 27 tasks add a clock-multiplexer or generated-clock wrapper. Their wrappers, specifications, and references do not come from the training-data generator.

\noindent\textbf{Construction and validation.}
Tasks based on public repositories retain their original RTL, and each pairs a specification with a reference SDC that loads cleanly in OpenSTA and whose exported state is fixed by a content hash. GPT-6-Astra~\cite{openai2026gpt6astra} writes the specifications for all 140 benchmark tasks, in a format similar to SDCBench, whereas DeepSeek v4-pro generates those for training. Nearest-neighbor 5-gram Jaccard similarity between benchmark specifications and training data is 0.13 for general tasks and 0.14 for rare-type tasks.
To verify that the scorer detects incorrect constraints, we construct 386 negative controls by mutating reference SDCs, including incorrect clock periods, missing I/O delays, extra false paths, and constraints on nonexistent objects. All are rejected. A task passes only when the generated SDC loads without errors or warnings and its exported facts match the reference.

\begin{figure*}[t]
\begin{minipage}[t]{.675\textwidth}
\vspace{0pt}
\captionof{table}{Pass counts on SDCBench and \bench{} (metrics in \S\ref{sec:exp}). \sdcoder{}: R4 + injection.}
\label{tab:main}
\vspace{-3pt}
\centering
\scriptsize
\setlength{\tabcolsep}{1pt}
\begin{tabular*}{\linewidth}{@{\extracolsep{\fill}}lcccccccccccc@{}}
\toprule
& \multicolumn{6}{c}{SDCBench (22 tasks)}
& \multicolumn{6}{c}{\bench{} (140 tasks)} \\
\cmidrule(lr){2-7}\cmidrule(l){8-13}
& \multicolumn{2}{c}{Load $\uparrow$}
& \multicolumn{2}{c}{Semantic $\uparrow$}
& \multicolumn{2}{c}{False path $\uparrow$}
& \multicolumn{2}{c}{Load $\uparrow$}
& \multicolumn{2}{c}{Semantic $\uparrow$}
& \multicolumn{2}{c}{False path $\uparrow$} \\
\cmidrule(lr){2-3}\cmidrule(lr){4-5}\cmidrule(lr){6-7}
\cmidrule(lr){8-9}\cmidrule(lr){10-11}\cmidrule(l){12-13}
Method
& pass@1 & pass@5
& pass@1 & pass@5
& pass@1 & pass@5
& pass@1 & pass@5
& pass@1 & pass@5
& pass@1 & pass@5 \\
\midrule
\multicolumn{13}{@{}l}{\textit{General-Purpose Models}} \\
GPT-6.1-sol
& \textbf{19} & 19 & \textbf{19} & 19
& \textbf{19} & \textbf{19}
& \textbf{129} & 130 & 94 & 101 & 86 & 86 \\
GPT-6.1-terra
& 18 & 19 & 16 & 18 & 17 & \textbf{19}
& 121 & 129 & 88 & 96 & 86 & 87 \\
DeepSeek v4.1-flash
& 14 & 18 & 8 & 16 & 9 & 16
& 122 & \textbf{138} & 77 & 107 & 77 & 88 \\
DeepSeek v4-pro
& 14 & 18 & 13 & 16 & 12 & 16
& 95 & 112 & 32 & 52 & 24 & 35 \\
\midrule
\multicolumn{13}{@{}l}{\textit{RTL-Specialized Models}} \\
CodeV-R1-RL
& 0 & 2 & 0 & 1 & 0 & 2
& 0 & 7 & 0 & 0 & 4 & 20 \\
VeriReason
& 0 & 1 & 0 & 0 & 0 & 1
& 0 & 3 & 0 & 0 & 1 & 2 \\
\midrule
\multicolumn{13}{@{}l}{\textit{Open-Weight Code Models}} \\
Qwen2.5-Coder-14B
& 0 & 1 & 0 & 1 & 0 & 1
& 0 & 12 & 0 & 1 & 0 & 6 \\
Qwen3-Coder-30B
& 1 & 3 & 1 & 2 & 1 & 2
& 0 & 1 & 0 & 0 & 2 & 11 \\
\midrule
\multicolumn{13}{@{}l}{\textit{Ours}} \\
SFT model
& 16 & 19 & 13 & 16 & 14 & 18
& 92 & 101 & 81 & 85 & 70 & 78 \\
GRPO (R4)
& 17 & 19 & 14 & 16 & 16 & 18
& 99 & 102 & 88 & 91 & 73 & 77 \\
\sdcoder{}
& \textbf{19} & \textbf{21}
& \textbf{19} & \textbf{20}
& \textbf{19} & \textbf{19}
& 128 & 130
& \textbf{126} & \textbf{127}
& \textbf{89} & \textbf{90} \\
\bottomrule
\end{tabular*}
\end{minipage}
\hfill
\begin{minipage}[t]{.31\textwidth}
\vspace{0pt}
\centering
\includegraphics[width=\linewidth]{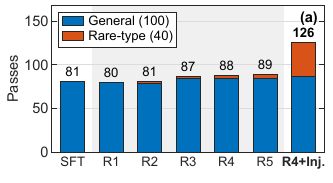}

\vspace{-1pt}
\includegraphics[width=\linewidth]{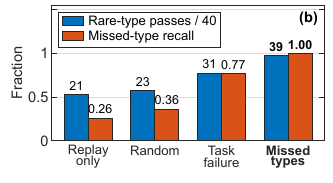}\par\vspace{-1pt}
\captionof{figure}{(a) \bench{} semantic passes per stage (shaded: GRPO; Inj.: injection). \mbox{(b) Selection:} 850 added examples, fixed replay.}
\label{fig:regmap}\label{fig:gran}
\end{minipage}
\end{figure*}

\section{Experiments}
\label{sec:exp}

\subsection{Setup}

\textbf{Benchmarks.} We evaluate on SDCBench~\cite{han2026llm4sdc} (22 tasks) and \bench{} (100 general and 40 rare-type tasks). A \emph{load pass} requires an SDC script to execute in OpenSTA without errors or warnings. A \emph{semantic pass} requires the exported constraint state to match the reference on \bench{}. Because SDCBench has no reference SDCs, we use the RTL clocking checks of \S\ref{sec:rtl}. A \emph{false-path pass} requires reference-state agreement on the 92 \bench{} tasks containing false paths, or clean loading with no RTL-rule violations for false paths or clock groups on SDCBench. Pass@1~\cite{chen2021codex} uses greedy decoding. Pass@5 counts a task as passed if any of five samples succeeds, using temperature 0.8 and top-$p$ 0.95, except for GPT-6.1 models (API default).

\textbf{Baselines.} We compare Qwen2.5-Coder-14B-Instruct~\cite{hui2024qwen25coder}, Qwen3-Coder-30B~\cite{yang2025qwen3}, CodeV-R1-RL~\cite{zhu2025codevr1}, VeriReason~\cite{wang2025verireason}, GPT-6.1-sol~\cite{openai2026gpt61sol}, GPT-6.1-terra~\cite{openai2026gpt61terra}, DeepSeek v4-pro~\cite{deepseek2026v4}, and DeepSeek v4.1-flash~\cite{deepseek2026v41flash}. All outputs use the same scoring; a setup/hold false-path pair on the same paths counts as one false path. Local-model pass@1 uses no repair or retry.

\textbf{Training.} We fine-tune Qwen2.5-Coder-14B-Instruct~\cite{hui2024qwen25coder} with rank-16 low-rank adaptation (LoRA)~\cite{hu2022lora} on the split of \S\ref{sec:coaug}. GRPO samples $G=4$ candidates per prompt with learning rate $10^{-6}$ and KL coefficient $\beta=0.02$.
We select checkpoints using CRG tasks sharing no module with SDCBench: 76 with reference SDCs and 33 without. The 33 reference-free tasks are evaluated with RTL clocking checks on their original specifications (\emph{original validation}) and versions restated from timing contracts (\emph{restated validation}). Neither SDCBench nor \bench{} is used for selection. GRPO retains the best-scoring checkpoint over five rounds (R1--R5) with increasing data coverage. Injection starts from R4 and retains the latest checkpoint whose pass count is within one task of R4 on each of the three validation sets.

\textbf{Research questions.} We evaluate overall performance (RQ1, \S\ref{sec:main}), identify what execution-based RL learns and misses (RQ2, \S\ref{sec:reach}), and test whether type-routed reference injection addresses those failures (RQ3, \S\ref{sec:routing}).

\subsection{Main Results (RQ1)}
\label{sec:main}

\textbf{General performance.} Table~\ref{tab:main} shows the pass rates of baselines and \sdcoder{} on different tasks. On the 100 general tasks, \sdcoder{} passes 87, compared with 84 for R4 and 72 for GPT-6.1-sol.
On the 40 rare-type tasks, it passes 39, compared with 4 for R4 and 22 for GPT-6.1-sol. Together, \sdcoder{} passes 126 of 140 tasks.
On SDCBench, injection increases semantic pass@1 from 14 to 19 of 22 tasks and pass@5 from 16 to 20. Load and false-path passes also improve.
Unlike \bench{}, SDCBench uses RTL-based semantic checks, which are also used in training.

\textbf{Evidence beyond targeted tasks.} Three checks support the rare-type result. First, these tasks were fixed before the diagnosis and are disjoint from training (\S\ref{sec:bench}). Second, under the same replay set and injection budget, random selection passes only 23 rare-type tasks (\S\ref{sec:routing}). Third, the general tasks contain 13 naturally occurring rare constraint statements on unchanged RTL (\S\ref{sec:bench}). \sdcoder{} reproduces 9, versus 2 for R4, including 4 of the 6 multicycle paths R4 misses.

\textbf{Failure analysis.} We compare failing scripts against their references using OpenSTA setup and hold slack at register data pins and outputs. Across the 140 \bench{} tasks, failures that remove or relax timing checks decrease from 18 for SFT to 12 for R4 and 6 after injection; such failures are the most harmful, because they hide timing violations rather than report them. Scripts causing OpenSTA to stop with an error decrease from 32 and 31 to 0. On rare-type tasks, 10 GPT-6.1-sol scripts contain invalid clock-latency commands, which we correct before scoring. After these corrections, 18 of the 40 GPT-6.1-sol scripts execute but fail semantic checks. Their errors include incorrect false paths, over-broad multicycle paths, and a second input delay overwriting the first.

\subsection{What Execution-Based Rewards Teach (RQ2)}
\label{sec:reach}

\textbf{GRPO improvements.} Semantic passes on \bench{} rise from 81 for SFT to 89 at R5 (Fig.~\ref{fig:regmap}(a)). Reward variants give similar validation gains: command acceptance alone, acceptance with worst slack, and RTL rewards with pair consistency raise reference-SDC validation passes from 37 to 60--61. Removing pair consistency from R4 changes semantic passes by at most 3. Loading is not the main bottleneck: deleting OpenSTA-flagged statements lets every R5 script outside the rare-type subset load, but adds no semantic passes.

\textbf{Missed constraint types.} R4 passes only 16 of 287 diagnostic GRPO-pool examples involving difficult constraint patterns, including through-point exceptions, exceptions targeting pins, clock multiplexers, case analysis on bus signals, and multicycle paths. On the reference-SDC examples in the GRPO pool, the greedy type-level probe identifies multicycle paths (recall 0), clock groups (0.005), and through-point exceptions (0.048) as missed types. In 235 outputs, the model invents a command name for clock groups. Data frequency does not fully explain these failures: multicycle paths occur in 3.3\% of SFT examples and coexisting generated clocks in 1.1\%, yet their post-GRPO recalls are 0 and 0.19, respectively. For example, the rare-type task in Fig.~\ref{fig:example} requires a two-cycle setup multicycle path from a register\label{sec:case}. R4 generates the correct cycle count but starts the path at the write-enable input port rather than at the register; OpenSTA loads the script, but the exported state exposes the wrong start point. \sdcoder{} generates the reference.

\textbf{Additional GRPO.} From R4, we test 289 more GRPO steps on the injection examples of \S\ref{sec:routing} plus 304 prompt pairs from earlier GRPO training, and a fivefold learning rate (Table~\ref{tab:reach}). Both improve clock groups and coexisting generated clocks, while the higher rate also improves through-point exceptions. However, multicycle-path recall remains 0 at every checkpoint in both runs; the higher rate also reduces restated validation passes from 32 to 0. In a separate sampling probe at temperature 1.0, 64 samples per relevant rare-type task yield no correct multicycle paths in 512 samples and only 6 correct clock groups in 1,856 samples. The latter gives $q_k\approx0.003$, requiring roughly $G\approx210$ samples for even odds of observing a correct instance (Eq.~\ref{eq:group}). Temperatures 1.2 and 1.5 still produce no correct multicycle paths.

\begin{table}[t]
\caption{Extra GRPO from R4: correct statements per type.}
\label{tab:reach}
\vspace{-3pt}
\centering\footnotesize\setlength{\tabcolsep}{1.2pt}
\begin{tabular*}{\columnwidth}{@{\extracolsep{\fill}}lcccccc@{}}
\toprule
 & Multi- & Clock & Coexist. & Through- & Tasks & Restated \\
 & cycle (32) & groups (32) & clk (33) & point (12) & (40) & (33) \\
\midrule
R4 (start) & 0 & 0 & 1 & 1 & 4 & 32 \\
289 more steps & 0 & 13 & 21 & 2 & 4 & 29 \\
$5\times$ learning rate & 0 & 16 & 26 & 10 & 9 & 0 \\
\bottomrule
\end{tabular*}
\end{table}

\begin{figure}[t]
\centering
\includegraphics[width=\columnwidth]{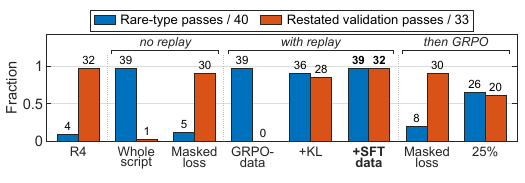}\par\vspace{-8pt}
\caption{Injection variants from R4: rare-type and restated validation passes.}
\vspace{-8pt}
\label{fig:route}
\end{figure}

\subsection{Type-Routed Injection (RQ3)}
\label{sec:routing}

For the example-selection and whole-script/replay ablations, we start from R4 and train for two SFT epochs at learning rate $2\times10^{-5}$. We test example selection under a fixed budget (Fig.~\ref{fig:gran}(b)), whole-script supervision and replay (Fig.~\ref{fig:route}), and injection without prior GRPO (Table~\ref{tab:nogrpo}).

\textbf{Example selection.} Random, task-failure, and missed-type selection each add 850 examples to the same 1,193 replay examples; replay alone adds no new examples. Missed-type selection passes 39 rare-type tasks, versus 31 for task-failure selection and 23 for random selection (Fig.~\ref{fig:gran}(b)). Task-failure selection also reduces general-task passes from 84 to 71, whereas missed-type selection increases them to 87. Of the task-failure-selected examples, 518 contain a missed-type statement; the missed-type recall shown in Fig.~\ref{fig:gran}(b) is 0.77, versus 1.00 for missed-type selection. The additional gain over random and task-failure selection thus comes from covering the diagnosed types.

\textbf{Whole-script supervision and replay.} Without replay, whole-script injection passes 39 rare-type tasks but only 1 of 33 restated validation tasks (Fig.~\ref{fig:route}). Masking the loss to missed-type statements preserves 30 restated validation passes but recovers only 5 rare-type tasks; 17 of its 25 original-validation failures introduce an unrequested clock-group constraint. This supports supervising complete scripts, including constraints that should remain absent. Replay is needed to retain earlier capabilities. GRPO-data replay matches the whole-script rare-type pass count but replaces the required edge-derived generated clock with an independent primary clock on every restated specification. A KL penalty recovers most restated validation passes. Adding 600 SFT examples to the replay yields 39 rare-type passes and keeps each validation score within one task of R4.
Variant selection uses validation only; SFT-data replay is the only variant that meets the checkpoint rule of \S\ref{sec:exp}. Rare-type passes, reported for all variants, were not used for selection.

\textbf{Injection without GRPO.} Table~\ref{tab:nogrpo} compares injection from R4 with injection directly from SFT, using either the examples selected for R4 or an SFT-specific type diagnosis. Both SFT-started variants recover most rare-type tasks but reduce general-task passes from the SFT baseline of 81 to 74 and 73. Type-balanced SFT on rare-type examples without GRPO or GRPO-derived replay passes 82 general tasks, but drops SDCBench from 13 to 6 and passes only 11 of 33 original validation tasks. Starting from R4 raises general-task passes from 84 to 87.

\begin{table}[t]
\caption{Injection start point: semantic pass@1 (shaded: \sdcoder{}).}
\label{tab:nogrpo}
\vspace{-3pt}
\centering\footnotesize\setlength{\tabcolsep}{1.5pt}
\begin{tabularx}{\columnwidth}{llc*{3}{>{\centering\arraybackslash}X}}
\toprule
Start & Injected examples & SDCBench & Rare-type & General & Total \\
 & & (22) & (40) & (100) & (140) \\
\midrule
SFT & none & 13 & 0 & 81 & 81 \\
SFT & selected for R4 & 15 & 38 & 74 & 112 \\
SFT & own diagnosis & 14 & 37 & 73 & 110 \\
SFT & rare types, no GRPO data & 6 & 38 & 82 & 120 \\
\rowcolor{blockfill} R4 & selected for R4 & \textbf{19} & \textbf{39} & \textbf{87} & \textbf{126} \\
\bottomrule
\end{tabularx}
\end{table}

\section{Conclusion}
\label{sec:conclusion}

We presented \sdcoder{}, a locally deployable 14B model for SDC generation. We find that execution-based GRPO improves overall correctness but struggles with constraint types it rarely generates correctly, particularly multicycle paths. To address this, we use constraint-level recall to identify missed types and fine-tune on complete reference scripts containing them, with replay to preserve previously learned capabilities. On \bench{}, \sdcoder{} passes 87 of 100 general tasks and 39 of 40 rare-type tasks, compared with 84 and 4 after GRPO alone and 72 and 22 for GPT-6.1-sol. It also matches GPT-6.1-sol on SDCBench at pass@1 (19/22). 

\bibliographystyle{IEEEtran}
\bibliography{refs}

\end{document}